\documentclass[journal,twocolumn]{IEEEtran}

\IEEEoverridecommandlockouts                              
\usepackage[utf8]{inputenc}
\usepackage{cite}
\usepackage{subcaption}
\usepackage{graphicx}
\usepackage{amsmath}
\usepackage{amssymb}
\usepackage[normalem]{ulem}
\usepackage[ruled,vlined]{algorithm2e}
\usepackage{framed}
\usepackage[table]{xcolor}
\usepackage{array}
\usepackage{booktabs, tabularx}
\usepackage{multirow}
\usepackage{bigints}
\usepackage{times}

\usepackage[]{algorithm2e}
\usepackage{bm}  % bold math symbols
\usepackage{color}
\usepackage{ulem}
\graphicspath{ {images/} }

\usepackage{balance}

\newcommand{\yudel}[1]{}
\usepackage{float}
\usepackage{caption}
\usepackage{subcaption}

\title{ Privacy-Preserving Authentication \\
for Secure Military 5G Networks}
\author{
\IEEEauthorblockN{Isabella D. Lutz, Ava Hill, and Matthew C. Valenti} \\
West Virginia University, Morgantown, WV, USA. \\
\vspace{-0.5cm}
}

\begin{document}

\maketitle
\thispagestyle{empty}
\pagestyle{empty}

%%%%%%%%%%%%%%%%%%%%%%%%%%%%%%%%%%%%%%%%%%%%%%%%%%%%%%%%%%%%%%%%%%%%%%%%%%%%%%%%
\begin{abstract}
As 5G networks gain traction in defense applications, ensuring the privacy and integrity of the Authentication and Key Agreement (AKA) protocol is critical. While 5G AKA improves upon previous generations by concealing subscriber identities, it remains vulnerable to replay-based synchronization and linkability threats under realistic adversary models. This paper provides a unified analysis of the standardized 5G AKA flow, identifying several vulnerabilities and highlighting how each exploits protocol behavior to compromise user privacy. To address these risks, we present five lightweight mitigation strategies.  We demonstrate through prototype implementation and testing that these enhancements strengthen resilience against  linkability attacks with minimal computational and signaling overhead.  Among the solutions studied, those introducing a UE-generated nonce emerge as the most promising, effectively neutralizing the identified tracking and correlation attacks with negligible additional overhead.   Integrating this extension as an optional feature to the standard 5G AKA protocol offers a backward-compatible, low-overhead path toward a more privacy-preserving authentication framework for both commercial and military 5G deployments.

\end{abstract}

\section{Introduction}
\label{intro}

Fifth-generation (5G) mobile networks are widely deployed and are increasingly being adopted in defense and national security applications \cite{elmasry,bhuyan}. Offering high data rates, low latency, and the ability to connect a large number of devices, 5G enables mission-critical capabilities such as real-time situational awareness, autonomous logistics, distributed command and control, and immersive training environments \cite{ Saarnisaari}. Recognizing this potential, the U.S. Department of Defense (DoD) has launched a range of programs to assess and adapt commercial 5G technologies for military use. Through testbeds at installations such as Hill Air Force Base and Joint Base Lewis-McChord, the DoD is evaluating 5G for smart warehousing, spectrum co-existence, and tactical communications \cite{mhl}. The establishment of a FutureG office further highlights DoD’s strategic interest in shaping secure and resilient next-generation wireless networks \cite{dod}.

The integration of 5G technology into military operations places heightened emphasis on the robustness of its underlying security protocols. At the center of 5G access security is the Authentication and Key Agreement (AKA) protocol, standardized in 3GPP TS 33.501 \cite{3GPP.TS.33.501}. This protocol enables mutual authentication between user equipment (UE) and the network and derives fresh session keys to ensure the confidentiality and integrity of subsequent communications. A major improvement over earlier generations, such as LTE—where the UE’s identity was often transmitted in the clear—is the introduction of identity protection in 5G via public key encryption of the subscriber's permanent identifier (SUPI) \cite{koutsos}. Additionally, the use of sequence numbers helps mitigate the threat of replay attacks by ensuring message freshness.

Despite these enhancements, recent research has shown that the 5G AKA protocol remains vulnerable to certain types of exploitation \cite{koutsos,fouque,cheng2022tracking}. Specifically, attackers can still correlate concealed identities across sessions—even when encrypted—or replay old messages to trigger resynchronization procedures that leak sensitive information. These vulnerabilities compromise unlinkability and can allow adversaries to track a UE's location over time. This is particularly concerning in military environments, where UEs may move frequently between cells and operate under persistent threat of surveillance. The reliance on sequence numbers without cryptographic proof of freshness from the UE side leaves the protocol susceptible to linkability and replay-based attacks. As a result, some of the privacy protections that 5G was intended to provide remain incomplete in practice.

% The fifth-generation (5G) mobile network architecture brings unprecedented performance in terms of throughput, latency, and device density, enabling a multitude of novel applications ranging from ultra-reliable low-latency communication (URLLC) to massive machine-type communication (mMTC). At the heart of 5G security lies the Authentication and Key Agreement (AKA) protocol, as specified in 3GPP TS 33.501, which must simultaneously protect subscriber identity, ensure mutual authentication between the User Equipment (UE) and the network, and derive fresh session keys for confidentiality and integrity protection. To achieve these goals, 5G AKA leverages SUCI-based SUPI concealment for subscriber privacy, an Elliptic-Curve Diffie–Hellman exchange over Curve25519 (X25519) for key establishment, and a layered key-derivation hierarchy built on HKDF/HMAC-SHA-256.

% Despite these robust design choices, recent studies have demonstrated that certain aspects of 5G AKA—particularly its SQN-based freshness mechanism and SUPI linkage resistance—remain susceptible to replay and linkability attacks under realistic adversary models. An attacker capable of observing protocol runs or injecting old messages can exploit the sequence-number synchronization checks to force resynchronization procedures or correlate SUCI values across sessions, undermining subscriber privacy. Furthermore, the existing protocol does not provide explicit cryptographic proof of the 'freshness' originating from the UE side, leaving gaps in the guarantees of the session key agreement.

This paper makes three contributions to the study of privacy in the 5G Authentication and Key Agreement (AKA) protocol. First, it presents a unified and precise exposition of three privacy vulnerabilities using consistent terminology and notation to enable direct comparison. Second, it evaluates five mitigation strategies, some drawn from existing literature and others newly proposed or refined, assessing their effectiveness under realistic communication and computation constraints. Third, it introduces a complete Python implementation of the 5G AKA protocol and its variants, which is used to validate the proposed solutions through reproducible testing scenarios. The results highlight the benefits of introducing lightweight and privacy-preserving enhancements to 5G AKA, especially in the context of military communications.

% In this paper, we present a two‐fold contribution. First, we perform a security analysis of the standardized 5G‐AKA flow, identifying both synchronization and linkability weaknesses. Second, we propose a lightweight UE‐side challenge extension: as soon as the UE has established its fresh encryption and integrity context from the network’s initial challenge, it immediately generates an eight‐byte random nonce, wraps it in a new NAS Information Element, and sends it using those same protections. This adds no extra round trips—no additional messages beyond those already secured by NAS security—but forces the network to prove that its sequence number is in sync with the UE’s before the UE ever processes the authentication token.

% The paper is organized as follows. Sec. II provides an overview of the 5G AKA protocol. Sec. III identifies three security vulnerabilities in the 5G AKA protocol that compromise unlinkability. Sec. IV proposes several improvements to 5G AKA which serve as a countermeasure to the previously described linkability attacks.  Sec V describes our implementation and testing of the protocol. Finally, Sec. VI concludes the paper.

\section{Overview of 5G AKA}
\label{section2}
\label{sec:background}

This section presents an overview of the 5G AKA protocol, as standardized in 3GPP TS 33.501 \cite{3GPP.TS.33.501}, emphasizing the components and procedures most pertinent to the security vulnerabilities discussed in Section III and the proposed mitigations in Section IV.  

Note: in the following discussion, we use the notation 
% $\langle\cdot,\cdot\rangle$ to denote a tuple,
$\mathsf{a} || \mathsf{b}$ to denote concatenation of the bitstreams $\mathsf{a}$ and $\mathsf{b}$.  Additionally, we use $\mathsf{a} \oplus \mathsf{b}$ to denote the bitwise exclusive-or of the 
%the  %DUPLICATED THE
same-length bitstreams $\mathsf{a}$ and $\mathsf{b}$.

% We begin by describing the core network entities involved in the authentication process, including their roles and responsibilities. We then detail the sequence of messages exchanged between the User Equipment (UE), Serving Network (SN), and Home Network (HN) during a standard AKA procedure.

% This section provides an overview of the 5G Authentication and Key Agreement (AKA) protocol, with a focus on the aspects most relevant to the security vulnerabilities discussed in Section III and the proposed mitigations in Section IV. We begin by introducing the key components of the 5G core network that participate in the authentication process. We then outline the sequence of messages exchanged during AKA.
% and describe the key derivation steps that take place immediately following successful authentication.

\subsection{5G Core Architecture}
\label{sec:background-arch}

% The 5G AKA protocol involves five principal logical elements of the 5G Core Network (see Fig.~1):

% \begin{figure}[t]
% \centering
% \includegraphics[width=0.43\textwidth]{images/Fig1}
% \caption{{\small Simplified 5G network architecture.}}
% \vspace{-0.25cm}
% \label{fig:net}
% \end{figure}

The 5G AKA protocol involves three primary participants: the \emph{User Equipment} (UE), the \emph{Home Network} (HN), and the \emph{Serving Network} (SN). While each of these entities may comprise multiple internal components, their internal structure is not essential for the discussion that follows.

\paragraph{User Equipment (UE)}  
The UE is the subscriber’s device and contains a secure hardware element—typically a \emph{Universal Subscriber Identity Module} (USIM)—responsible for storing sensitive credentials and performing cryptographic operations. The USIM maintains the following:
\begin{itemize}
\item The permanent identifier of the subscriber, known as the \emph{Subscription Permanent Identifier} (SUPI).
\item A long-term shared secret key $\mathsf{K}$, known only to the UE and its Home Network.
\item A locally maintained sequence number, $\mathsf{SQN}_\mathsf{UE}$, used for freshness checks.
\end{itemize}
During the authentication process, the UE initiates communication by sending a concealed version of the SUPI, known as the \emph{Subscription Concealed Identifier} (SUCI). It then verifies a challenge issued by the network and, if necessary, initiates a resynchronization procedure in the event of sequence number mismatch.

\paragraph{Home Network (HN)}
The HN manages subscriber identity and authentication credentials. For each registered UE, the HN maintains:
\begin{itemize}
\item The SUPI associated with the subscriber.
\item The corresponding long-term shared secret key $\mathsf{K}$.
\item Its own sequence number counter $\mathsf{SQN}_\mathsf{HN}$, which may differ from the UE’s.
\end{itemize}
The HN is responsible for decrypting the received SUCI to recover the SUPI, retrieving the appropriate credentials, and generating the authentication challenge. It also computes the expected response and provides a cryptographic hash of this value (denoted $\mathsf{HXRES}^*$) to the SN.

\paragraph{Serving Network (SN)}
The SN is the access network to which the UE is currently connected. The UE connects wirelessly to the SN through a gNB (next-generation NodeB) residing in the SN. The SN does not possess the subscriber’s long-term key $\mathsf{K}$ or the sequence number $\mathsf{SQN}$. Instead, it acts as an intermediary, relaying AKA messages between the UE and the HN. After receiving the UE’s response to the challenge, the SN verifies authentication success by comparing the hashed response it computes from the UE's message to the expected hash $\mathsf{HXRES}^*$ previously provided by the HN.

\subsection{5G AKA Message Sequence}
\label{sec:background-flow}
\begin{figure}[t]
\centering
\includegraphics[width=0.325\textwidth]{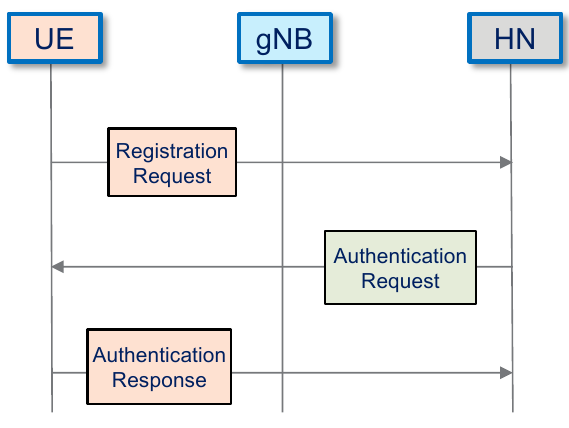}
\caption{{\small Messages used by the AKA protocol.}}
\vspace{-0.5cm}
\label{message:fig}
\end{figure}

Fig. \ref{message:fig} illustrates the sequence of messages exchanged between the UE and the HN, relayed through the gNB of the SN, during a successful execution of the AKA procedure.

% Before describing each step, we first introduce the cryptographic primitives used and the notation adopted. 

The cryptographic primitives are those specified for 5G AKA. In particular, they include the keyed symmetric functions $f_1$, $f_2$, $f_5$, $f_{1,\ast}$, and $f_{5,\ast}$. In the description that follows, these functions are all keyed by the shared secret $\mathsf{K}$, which is securely stored at the UE and HN.  In addition, asymmetric public-key cryptography with ephemeral key encapsulation is used to encrypt the SUPI. This involves both a public key $\mathsf{p}_\mathsf{HN}$ advertised by the HN and an ephemeral secret key $\mathsf{s}_\mathsf{UE}$ generated by the UE during each authentication attempt. We use the notation $\left.\{ \cdot \}\right|_\mathsf{p}^\mathsf{s}$ to denote such an encryption using public key $\mathsf{p}$ and secret key $\mathsf{s}$. 

% The AKA protocol relies on sequence numbers, which are separately stored at the UE and the HN: The version stored at the UE is denoted $\mathsf{SQN}_\mathsf{UE}$ and the version stored at the HN is $\mathsf{SQN}_\mathsf{HN}$. 

\paragraph{Registration Request}
When the UE does not have a valid non-access stratum (NAS) security context, it initiates the AKA protocol by sending a \texttt{Registration Request} message that includes a SUCI, which is a concealed version of its SUPI. For 3GPP access, the SUPI is typically an IMSI (International Mobile Subscriber Identity), composed of three fields: the Mobile Country Code (MCC), the Mobile Network Code (MNC), and the Mobile Subscription Identification Number (MSIN).

The SUCI, which is carried in the \texttt{5GS Mobile Identity} information element, includes the MCC and MNC in plaintext but protects the MSIN using the Elliptic Curve Integrated Encryption Scheme (ECIES). 
Under ECEIS \emph{Profile A}, which is the most commonly used of the profiles defined by 3GPP, the UE constructs the SUCI by retrieving 
%Among the ECIES profiles defined by 3GPP, Profile A is the one most widely adopted in 5G NR deployments and the one used in this implementation. To construct the SUCI, the UE retrieves 
the HN’s public key $\mathsf{p}_\mathsf{HN}$ and generates its own ephemeral key pair $\langle \mathsf{p}_\mathsf{UE}, \mathsf{s}_\mathsf{UE}\rangle$. A shared secret is then computed as:
\begin{eqnarray}
\mathsf{Z} & = & \mathsf{X25519}(\mathsf{s}_\mathsf{UE}, \mathsf{p}_\mathsf{HN})
\end{eqnarray}
using the Diffie–Hellman function on Curve25519 as specified in RFC~7748. This shared secret is expanded with the ANSI~X9.63 KDF and SHA-256 to derive symmetric keys: one for AES-128 in counter mode (AES-CTR) to encrypt the MSIN and another for HMAC-SHA-256 to provide integrity protection. We denote the resulting SUCI construction as:
\begin{eqnarray}
\mathsf{SUCI} & = & \left. \left\{ \mathsf{MSIN} \right\} \middle|^{\mathsf{s}_\mathsf{UE}}_{\mathsf{p}_\mathsf{HN}}
\right.
\end{eqnarray}
The final \texttt{SUCI} information element therefore includes the ciphertext, the UE’s ephemeral public key $\mathsf{p}_\mathsf{UE}$, and a MAC for integrity protection.

% To conceal its permanent identifier (SUPI), the UE first retrieves the Home Network’s public key (via broadcast system information or dedicated provisioning) and constructs a Subscription Concealed Identifier (SUCI) as follows:
% \[
%   \underbrace{\texttt{0x01}}_{\text{Scheme ID}}
%   \;\|\;
%   \underbrace{\texttt{HNID}}_{\text{Home Network ID}}
%   \;\|\;
%   \underbrace{\mathsf{Enc}_{\mathsf{HN\_pub}}(\mathsf{MSIN})}_{\text{Encrypted MSIN}}
% \]
% Here, \(\mathsf{Enc}_{\mathsf{HN\_pub}}\) denotes ECIES encryption of the Mobile Subscriber Identification Number under the network’s public key, per 3GPP TS 33.501 \cite{3gpp33501}. The UE then sends:
% \[
%   \mathsf{SUCI}
%   \;\|\;
%   \texttt{KSI} = \text{null}
% \]
% in its Registration Request. Upon receipt, the AMF forwards the SUCI to the AUSF, which invokes the ARPF’s SIDF to decrypt SUCI → SUPI and generate a fresh Authentication Vector. Only after the AV exchange completes do the UE and network derive and install NAS integrity and cipher keys (\(\mathsf{IK_{NAS}},\mathsf{CK_{NAS}}\)), ensuring that the SUPI never appears in cleartext in any subsequent NAS messages. 

\paragraph{Authentication Request}
In the HN, the SUCI is decrypted using the HN's secret key $\mathsf{s}_\mathsf{HN}$ and the UE's ephemeral public key $\mathsf{p}_\mathsf{UE}$, revealing the MSIN.  This identifier is used to retrieve the long-term shared secret $\mathsf{K}$ associated with the identified subscriber.  

The HN then constructs a challenge, comprised of the following elements:
\begin{itemize}
\item A 128-bit random nonce $\mathsf{RAND}$, freshly generated for this session.
\item A 128-bit anonymity key $\mathsf{AK}$ used to conceal $\mathsf{SQN}_\mathsf{HN}$, derived as: $\mathsf{AK} = f_5(\mathsf{RAND})$.
\item The \emph{concealed sequence number} $\mathsf{CONC} = \mathsf{SQN}_\mathsf{HN} \oplus \mathsf{AK}$.
\item An \emph{Authentication Management Field} (AMF) which is used by the operator to encode algorithm preferences, protocol versioning, and other policy-related information.
\item A \emph{message authentication code} (MAC) computed as 
$\mathsf{MAC} = f_1(\mathsf{SQN_{HN}}\|\mathsf{RAND}\|\mathsf{AMF})$.
\item An \emph{Authentication Token} (AUTN) constructed as
$\mathsf{AUTN} = \mathsf{CONC} 
\| \mathsf{AMF}
\| \mathsf{MAC}
$.
\end{itemize}
The challenge, consisting of only $\mathsf{RAND}$ and $\mathsf{AUTN}$, is encapsulated in the \texttt{Authentication Request} message sent from the HN to the UE via the SN.  In parallel, the HN also computes the hash of the expected response, denoted $\mathsf{HXRES}^*$, and forwards it to the SN for later verification against the UE’s actual response.

\paragraph{Authentication Response}
Upon reception of $\mathsf{RAND}$ and $\mathsf{AUTN}$, the UE performs the following calculations:
\begin{enumerate}
\item Using the received $\mathsf{RAND}$ and its knowledge of $\mathsf{K}$, it computes $\mathsf{AK}'=f_5(\mathsf{RAND})$.  Note that $\mathsf{AK}' = \mathsf{AK}$ iff the HN and UE use the same key.
\item Using $\mathsf{AK}'$ and the masked sequence number contained in $\mathsf{AUTN}$, it demasks the sequence number as 
$\mathsf{SQN}' =  (\mathsf{SQN}_\mathsf{HN} \oplus \mathsf{AK}) \oplus \mathsf{AK}'$.  When the UE and HN use the same keys $\mathsf{SQN}' = \mathsf{SQN}_\mathsf{HN}$.
\item Using $\mathsf{SQN}'$, $\mathsf{RAND}$, the $\mathsf{AMF}$ contained in $\mathsf{AUTN}$, and its knowledge of $\mathsf{K}$, the UE computes an estimate of the MAC as $\mathsf{MAC}' = f_1(\mathsf{SQN}' \| \mathsf{RAND} \| \mathsf{AMF})$.   If the computed $\mathsf{MAC}'$ does not match the received $\mathsf{MAC}'$, then a \texttt{mac-failure} is flagged because the UE and HN do not possess the same $K$. 
\item The unmasked $\mathsf{SQN}'$ is checked for freshness by making sure it is in the window $\mathsf{SQN}_\mathsf{UE} < \mathsf{SQN}' \leq \mathsf{SQN}_\mathsf{UE} +\mathsf{W}$, where $\mathsf{W}$ is the size of an \emph{acceptance window}, which is a value stored in the USIM or provisioned by the operator (typically 32 or 64).  If it is outside this window then a \texttt{synch-failure} is flagged.
\item Assuming that there is no failure, the UE computes its response by using $\mathsf{RAND}$ and its knowledge of $\mathsf{K}$ as $\mathsf{RES} = f_2( \mathsf{RAND} )$.  
\item The UE finally computes the \emph{extended response} $\mathsf{RES}*$ by using a key distribution function (KDF) with binding inputs $\mathsf{RES}$, $\mathsf{RAND}$, and the serving network identity.
\end{enumerate}
The UE sends $\mathsf{RES}*$ to the SN within the \texttt{Authentication Response} message.  The SN then compares a hash of the received $\mathsf{RES}*$ against $\mathsf{HXRES}^*$, which had been previously sent by the SN.  If the two hashes match, the SN concludes that the UE has been successfully authenticated.

\paragraph{Authentication Failure}
The UE may encounter one of two kinds of failures, a \texttt{mac-failure}, which occurs when the UE and network possess different keys $\mathsf{K}$, or a \texttt{synch-failure}, which arises when the received SQN is outside the expected range.  In either case, the UE responds with an \texttt{Authentication Failure} message, setting the \texttt{Cause Value} information element that it contains to indicate the type of failure encountered.

In the case of a \texttt{mac-failure} the authentication is aborted, and the UE must either issue a new Registration Request or attempt to connect to a different gNB.  On the other hand, in the case of \texttt{synch-failure}, a resynchronization procedure is initiated.  This involves the UE computing an \emph{Authentication Synchronization Token} $\mathsf{AUTS}$, defined as:
\begin{eqnarray}
\mathsf{AUTS}
& = &
\mathsf{CONC}
\|
\mathsf{MAC\text{-}S}
\end{eqnarray}
where the \emph{concealed sequence number} is:
\begin{eqnarray}
\label{conc}
\mathsf{CONC}
& = &
\mathsf{SQN}_\mathsf{UE}
\oplus
f_{5,*} ( \mathsf{RAND} )
\end{eqnarray}
and the MAC is:
\begin{eqnarray}
\mathsf{MAC\text{-}S}
& = &
f_{1,*} ( \mathsf{SQN}_\mathsf{UE} \| \mathsf{RAND} ).
\end{eqnarray}
% Hence, the $\mathsf{AUTS}$ contains a covered version of the UE's current sequence number along with a MAC that depends on both the sequence number and the received RAND.  

Upon receiving $\mathsf{AUTS}$, the HN can retrieve 
$\mathsf{SQN}_\mathsf{UE}$ and use it to set its own counter
accordingly: $\mathsf{SQN}_\mathsf{HN} = \mathsf{SQN}_\mathsf{UE}$.  The network then sends a fresh authentication vector by issuing a new \texttt{Authentication Request} message.   Assuming no subsequent synchronization error, the UE will then respond with a valid \texttt{Authentication Response}.

\section{Security Vulnerabilities in 5G AKA}
\label{section3}

% \emph{Bella -- I plan to revise this to cover the Failure Message Attack and Encrypted IMSI replay attack of Koutsos.  I'm not sure what other kinds of attacks need to be described to set up your solution.}

% \subsection{Linkability Attacks}
% \label{sec:sync-attacks}

Having established the necessary background on the AKA protocol, we now examine several of its shortcomings.  We focus on vulnerabilities that can be exploited to compromise the privacy of the UE.  Although 5G AKA introduces significant improvements over earlier generations -- most notably through the use of public-key encryption to conceal the SUPI and hence prevent IMSI catcher attacks \cite{strobel} -- certain operational behaviors described herein still expose the UE to tracking and linkability attacks.

Our analysis centers on protocol-level exploits that adversaries can leverage to correlate messages across sessions, trigger distinguishable responses, or infer subscriber activity based on observable message patterns. These attacks do not require compromising cryptographic primitives but instead take advantage of subtle design choices, such as the use of sequence numbers for freshness and the UE’s behavior when responding to authentication failures. The resulting privacy leaks can enable location tracking, persistent identity correlation, and denial-of-service vectors, which are particularly concerning in adversarial or high-mobility environments.

In the remainder of this section, we describe these vulnerabilities in detail, categorize the types of attacks they enable, and highlight their implications for future network design.

\subsection{The Failure Message Attack}

The \emph{Failure Message Attack}, first described by Arapinis et al. \cite{arapinis}, demonstrates that, even when a UE's permanent identity is concealed, privacy can still be compromised through protocol behavior. The attack exploits the fact that the \texttt{Authentication Failure} message is transmitted in unencrypted cleartext and includes a \texttt{Cause Value} indicating the reason for the failure.

To mount the attack, an adversary first passively captures a legitimate \texttt{Authentication Request} message exchanged between the home network (HN) and a victim UE with permanent identity $\mathsf{IMSI}$. This message is stored for later use. The attacker then initiates a new authentication session and replays the captured \texttt{Authentication Request} to another UE with identity $\mathsf{IMSI}'$, encountered at a later time. $\mathsf{IMSI}'$ may or may not correspond to the original target.

The outcome depends on whether $\mathsf{IMSI}'$ matches $\mathsf{IMSI}$. If $\mathsf{IMSI}' \neq \mathsf{IMSI}$, the message authentication code (MAC) will fail verification, and the UE will respond with a \texttt{Cause Value} indicating a \texttt{mac-failure}. If instead $\mathsf{IMSI}'=\mathsf{IMSI}$, the MAC will verify correctly, but the sequence number will appear out of sync, leading to a \texttt{synch-failure}. In both cases, the adversary receives the UE’s response and inspects the unencrypted \texttt{Cause Value}. A \texttt{mac-failure} reveals that the current UE is not the same as the original, whereas a \texttt{synch-failure} confirms a match.

Although this attack does not expose the UE's full identity, it violates unlinkability by enabling the adversary to determine whether two sessions involve the same subscriber. Repeated use of this technique at different locations enables device tracking, posing a significant threat to user privacy—especially in sensitive or high-security environments.

\subsection{The SUCI Replay Attack}

The \emph{SUCI Replay Attack}, first introduced by Fouque et al. \cite{fouque}, enables an adversary to compromise user privacy despite the concealment of the subscriber's permanent identity. The attack begins when the adversary captures a $\mathsf{SUCI}$ transmitted by a victim UE with identity $\mathsf{IMSI}$. Later, upon encountering another UE with identity $\mathsf{IMSI}'$, the adversary replaces that UE’s freshly generated $\mathsf{SUCI}'$ with the previously stored $\mathsf{SUCI}$. As a result, the network responds with an \texttt{Authentication Request} intended for $\mathsf{IMSI}$, containing a valid challenge and an in-window sequence number.

The response of the current UE depends on whether $\mathsf{IMSI}'$ matches $\mathsf{IMSI}$. If $\mathsf{IMSI}' \neq \mathsf{IMSI}$, the MAC verification fails, prompting the UE to reply with an \texttt{Authentication Failure}. Conversely, if $\mathsf{IMSI}' = \mathsf{IMSI}$, the MAC will validate successfully and—assuming the sequence number is in sync—the UE will return a valid \texttt{Authentication Response}. Thus, the attacker can infer whether two sessions involve the same subscriber simply by observing the type of message sent by the UE. An \texttt{Authentication Failure} indicates a mismatch; a valid \texttt{Authentication Response} confirms a match.

While this attack shares similarities with the \emph{Failure Message Attack}, it differs in a critical respect: the attacker does not need to inspect the \texttt{Cause Value} in a failure message. Instead, it only needs to distinguish between the message types. This distinction renders ineffective countermeasures such as encrypting \texttt{Authentication Failure} messages (see IV.A below). Like its predecessor, the SUCI Replay Attack poses a serious threat to user privacy by allowing persistent tracking and correlation of subscriber activity.

\subsection{The AUTS Attack}

An attack targeting the resynchronization procedure was proposed by Cheng and Shen \cite{cheng2022tracking}, which we refer to as the \emph{AUTS Attack}. The objective is to infer the victim UE’s sequence number. Like the \emph{Failure Message Attack}, it begins with the adversary passively capturing a legitimate \texttt{Authentication Request} message directed to a UE with identity $\mathsf{IMSI}$. This message is subsequently replayed twice at two sufficiently different times to the same UE, each time triggering a synchronization failure and causing the UE to respond with an \texttt{Authentication Failure} message containing an \texttt{AUTS}.

The first \texttt{AUTS} contains a concealed sequence number $\mathsf{CONC}$ corresponding to the UE’s sequence number at that time, $\mathsf{SQN}_\mathsf{UE}$, while the second contains $\mathsf{CONC}'$ corresponding to a later sequence number, $\mathsf{SQN}_\mathsf{UE}'$. Because both \texttt{Authentication Request} messages use the same random challenge $\mathsf{RAND}$, the anonymity key $\mathsf{AK}^* = f_{5,\ast}(\mathsf{RAND})$ remains constant, and it follows from (\ref{conc}) that:
\begin{eqnarray}
\mathsf{CONC}
& = &
\mathsf{SQN}_\mathsf{UE}
\oplus
f_{5,\ast} ( \mathsf{RAND} ) \\
\mathsf{CONC}'
& = &
\mathsf{SQN}_\mathsf{UE}'
\oplus
f_{5,*} ( \mathsf{RAND} )
\end{eqnarray}
Taking the XOR of both expressions yields:
\begin{eqnarray}
\label{concplus}
\mathsf{CONC} \oplus \mathsf{CONC}'
& = &
\mathsf{SQN}_\mathsf{UE}
\oplus
\mathsf{SQN}_\mathsf{UE}'
\end{eqnarray}
as the anonymity key cancels itself. Using the method described in \cite{borgaonkar}, the attacker can then infer $\mathsf{SQN}_\mathsf{UE}$ from the differential $\mathsf{CONC} \oplus \mathsf{CONC}'$.

This attack enables tracking of the UE associated with identity $\mathsf{IMSI}$ by estimating the number of authentications it has performed, which serves as a proxy for dwell time and activity within a given area. Over time, this can reveal behavioral patterns and routines of the targeted UE.

Furthermore, as in the \emph{Failure Message Attack}, the attacker can confirm whether the victim UE matches the captured $\mathsf{IMSI}$ by examining the \texttt{Cause Value} in the \texttt{Authentication Failure} message.

\section{Proposed 5G AKA Improvements}
\label{section4}

Having outlined several security vulnerabilities in Section III, we now turn to potential solutions aimed at mitigating these issues. Multiple approaches are considered, each evaluated in terms of their associated tradeoffs. Unless otherwise stated, we focus on solutions that operate within the same design constraints as the 5G AKA protocol—namely, those that require the standard three-message exchange shown in Fig. 1, utilize the same cryptographic primitives, and do not increase the number of random values that must be generated at the UE. 

\subsection{Encrypted Failure Message}
The \emph{Failure Message Attack} exploits the fact that the UE issues different responses depending on the type of error encountered. Specifically, if a replayed challenge is presented to the previously encountered UE, it will set the \texttt{Cause Value} to \texttt{synch-failure} and initiate a resynchronization process. Conversely, if the challenge is presented to a different UE, then it sets the \texttt{Cause Value} to \texttt{mac-failure}. This behavioral discrepancy enables an attacker to correlate sessions and infer whether the same UE has been encountered, thereby compromising user privacy.

One proposed mitigation is to encrypt the \texttt{Authentication Failure} message, thereby preventing an observer from inspecting the \texttt{Cause Value} field. As early as 2012, Arapinis et al.~\cite{arapinis} proposed using the same public key employed for SUPI concealment to encrypt failure messages. However, for this to be effective, the ciphertexts must be padded or formatted to ensure that their lengths do not reveal the error type.

While this solution does not increase the number of messages exchanged, it does increase the computational cost by requiring an additional public-key encryption. In the event of either a MAC failure or an SQN error, the UE must now perform two public-key encryption operations.

\subsection{Encrypted Response}
While encrypting the failure message may counter basic forms of the \emph{Failure Message Attack}, Koutsos~\cite{koutsos} demonstrated that it is ineffective against the more advanced \emph{SUCI Replay Attack} described in Section~III.B. In this attack, the replayed authentication challenge includes a valid sequence number, so no resynchronization is triggered. Consequently, the UE either responds with an \texttt{Authentication Response} if it shares the same secret $\mathsf{K}$ as the HN, or returns an \texttt{Authentication Failure} otherwise.

Encrypting only the \texttt{Authentication Failure} message is insufficient, as an attacker can still distinguish between a successful and unsuccessful authentication attempt based on the type of message returned. To mitigate this, we propose enhancing the encryption scheme to include both the \texttt{Authentication Failure} and the \texttt{Authentication Response} messages. Furthermore, the resulting ciphertexts must be padded to a fixed, uniform length to prevent inference based on message size.

This approach prevents external observers from learning the authentication outcome by inspecting message type or length. However, it comes at the cost of requiring two public-key encryption operations for every authentication attempt, regardless of whether the outcome is success or failure.

\subsection{UE Transmission of SQN}
In the AKA protocol, the UE and HN each maintain an independent copy of SQN. When these values diverge, the UE initiates a resynchronization procedure. Unfortunately, the resulting resynchronization message exposes behavioral differences that can be exploited by adversaries, as demonstrated in the \emph{Failure Message Attack} and the \emph{AUTS Attack}.

A more robust alternative is to transmit the UE’s current SQN alongside its identity in the initial message. Specifically, the SQN can be encrypted together with the MSIN to form an enhanced SUCI:
\begin{eqnarray}
\mathsf{SUCI} & = & \left. \left\{ \mathsf{MSIN} \| \mathsf{SQN}_\mathsf{UE} \right\} \middle|^{\mathsf{s}_\mathsf{UE}}_{\mathsf{p}_\mathsf{HN}} \right.
\end{eqnarray}

This strategy, central to the AKA+ protocol proposed in~\cite{koutsos}, enables the HN to synchronize its SQN with the UE’s without requiring a separate resynchronization message. As a result, synchronization errors—and the privacy leaks they cause—are eliminated, effectively mitigating both the \emph{Failure Message} and \emph{AUTS} attacks.

Notably, this approach maintains the same number of communication steps and public-key encryptions as standard 5G AKA. The only overhead is the encryption and inclusion of the 48-bit SQN in the SUCI.

\subsection{Packaging a Nonce in the SUCI}
Sequence numbers have been a feature of the 3GPP AKA protocols since \emph{Release 99} of the UMTS standard \cite{3GPP.TS.33.102}, where they were originally introduced to avoid requiring each UE to generate a random number—a task then considered computationally expensive. However, with modern hardware capabilities, it is now reasonable to expect a UE to generate a fresh random number for each authentication attempt. For instance, the SUCI generation process already involves creating an ephemeral key pair, which includes fresh randomness.

As an alternative to the AKA+ protocol’s inclusion of the SQN into the SUCI, we propose replacing the sequence number with a UE-generated nonce, $\mathsf{RAND}_\mathsf{UE}$. The enhanced SUCI is then defined as:
\begin{eqnarray}
\mathsf{SUCI} & = & \left. \left\{ \mathsf{MSIN} \| \mathsf{RAND}_\mathsf{UE} \right\} \middle|^{\mathsf{s}_\mathsf{UE}}_{\mathsf{p}_\mathsf{HN}} \right.
\end{eqnarray}

The remainder of the protocol follows the same structure as standard AKA, with the nonce $\mathsf{RAND}_\mathsf{UE}$ used in place of SQN. This modification entirely eliminates the need for sequence numbers, avoiding synchronization errors and removing the need for resynchronization requests. As a result, the UE becomes immune to the Failure Message Attack and the AUTS Attack. 

While this approach maintains the same message flow as AKA, its main overhead is the UE's need to generate a fresh random nonce for each session.  However, given the resulting security benefits—such as eliminating synchronization errors and thwarting failure message attacks—this trade-off may be justified. This aligns with emerging proposals for 6G authentication protocols, such as the one introduced in \cite{DMRN}, which also abandons the use of sequence numbers. Nevertheless, the full protocol in \cite{DMRN} relies on key encapsulation mechanisms that exceed the complexity and communication constraints we aim to preserve.

\subsection{Packaging a Nonce in the AUTS}
The \emph{AUTS Attack} described in Section~III.C exploits the fact that the value of $\mathsf{SQN}_\mathsf{UE}$ reported back to the HN is masked using an anonymity key $\mathsf{AK}^*$ that remains constant whenever the same \texttt{Authentication Request} is replayed. This occurs because the value of $\mathsf{RAND}$ in the request is reused to derive $\mathsf{AK}^* = f_5(\mathsf{RAND})$.

A proposed mitigation, originally suggested in~\cite{cheng2022tracking}, is for the UE to generate its own random nonce $\mathsf{RAND}_\mathsf{UE}$ and use it to derive a fresh anonymity key:
\begin{eqnarray}
\mathsf{AK}^* & = & f_5\!\left(\mathsf{RAND}_\mathsf{UE}\right)
\end{eqnarray}
However, implementation details were not fully specified in~\cite{cheng2022tracking}. In particular, the value of $\mathsf{RAND}_\mathsf{UE}$ must be included in the \texttt{AUTS} message so that the HN can reconstruct $\mathsf{AK}^*$ and unmask the reported $\mathsf{SQN}_\mathsf{UE}$.

While this approach is similar in spirit to the use of a UE-generated nonce in the SUCI (as discussed in Section~IV.D), it has the advantage of requiring the nonce only during resynchronization. However, it does not eliminate the need for sequence numbers and remains susceptible to the \emph{Failure Message Attack}.

\section{Implementation and Testing}
\label{section5}
\subsection{Implementation Overview}
The procedural details of 5G AKA and its improvements have already been presented in Sections II and IV. 
Here, we focus on how those designs were realized in our implementation, which was developed 
in Python to ensure portability, transparency, and reproducibility. The implementation supports the 
message flows of 3GPP TS~33.501, with configurable modes for both standardized 5G AKA and the modified 
nonce-in-SUCI variant described in Section IV. To maintain clarity and accessibility, we restrict our code to Python’s standard libraries 
(e.g., \texttt{secrets} for secure random values, \texttt{hashlib} and \texttt{hmac} for hashing and integrity checks, and 
\texttt{struct} for byte encoding). All intermediate values---such as $\mathsf{AUTN}$ construction, MAC checks, and SUCI 
ciphertexts---are logged in a structured format to enable deterministic replay of experiments.  

We evaluated two replay scenarios in our testing, corresponding to the vulnerabilities analyzed in Section~III:  
(i) \textit{replay of an Authentication Request}, which underlies the Failure Message Attack of Section~III-A, and  
(ii) \textit{replay of a SUCI}, which relates to the SUCI Replay Attack of Section~III-B.  
In the baseline protocol, an adversary can exploit these cases to distinguish whether a subscriber is the same or different.  
In the nonce-in-SUCI design, however, Authentication Request replays are uniformly rejected, and SUCI re-use is detected and logged, shifting replay attempts into a form that higher-layer defenses can monitor.  

Table~\ref{tab:TableI} summarizes the observable outcomes of these replay experiments. For clarity, the table labels follow 5G AKA behavior. ``ok'' denotes a normal authentication success, where the UE accepts $\mathsf{AUTN}$ and returns $\mathsf{RES}^*$, which is then verified by the serving network. ``mac-failure'' denotes an integrity check failure, occurring when the UE cannot verify the MAC in $\mathsf{AUTN}$ (typically because the subscriber key does not match). ``synch-failure'' denotes a freshness failure, where the MAC is valid but the sequence number falls outside the UE’s acceptance window, leading the UE to return an $\mathsf{AUTS}$ (Authentication Token for Synchronization). Finally, in the nonce-in-SUCI design, replayed messages produce a \emph{uniform reject} outcome, meaning that the protocol always returns the same generic error response regardless of subscriber state. This uniform treatment prevents an adversary from distinguishing whether the replayed message was directed at the correct subscriber or not, thereby eliminating linkability. Logged nonce re-use can also be exploited by higher-layer systems (e.g., intrusion detection or anomaly monitoring).  

\subsection{Baseline 5G AKA Implementation}
We first implemented the baseline 5G AKA as specified in Section II, modeling all core entities of the system: the UE, the Serving Network (SN), and the Home Network (HN). This baseline implementation includes the essential protocol functions such as SUCI generation and parsing, $\mathsf{AUTN}$ construction and verification, sequence number management, and the resynchronization procedure, providing a faithful reproduction of the standardized 5G AKA flow. 

This implementation serves as both a functional reference model and a test environment in which replay-based and 
linkability attacks can be reproduced. When captured Authentication Requests or SUCIs are replayed, the baseline 
generates error responses whose type depends on the subscriber’s state. These differing responses 
are the root of the privacy vulnerabilities analyzed in Section~III.  

\subsection{Modified AKA with UE Nonce in SUCI}
We then implemented the nonce-in-SUCI design, introduced in Section IV.D. In this design, the UE 
generates a fresh random nonce and inserts it into the SUCI. On the home network side, we 
added a replay-detection step that maintains a cache of recently used nonces per subscriber. If a nonce is reused, the SUCI is rejected at the authentication layer and the event is logged as side information for higher-layer mitigations.  

This modification eliminates the need for SQN management and resynchronization. The $\mathsf{AUTN}$ no longer encodes 
a masked sequence number; instead, integrity protection is bound to the UE nonce. As a result, all replayed SUCIs 
are intercepted directly by the network and all replayed Authentication Requests produce uniform rejections, ensuring 
that no distinguishable error codes are exposed to an adversary.

\begin{table}[t]
  \centering
  \includegraphics[width=\linewidth]{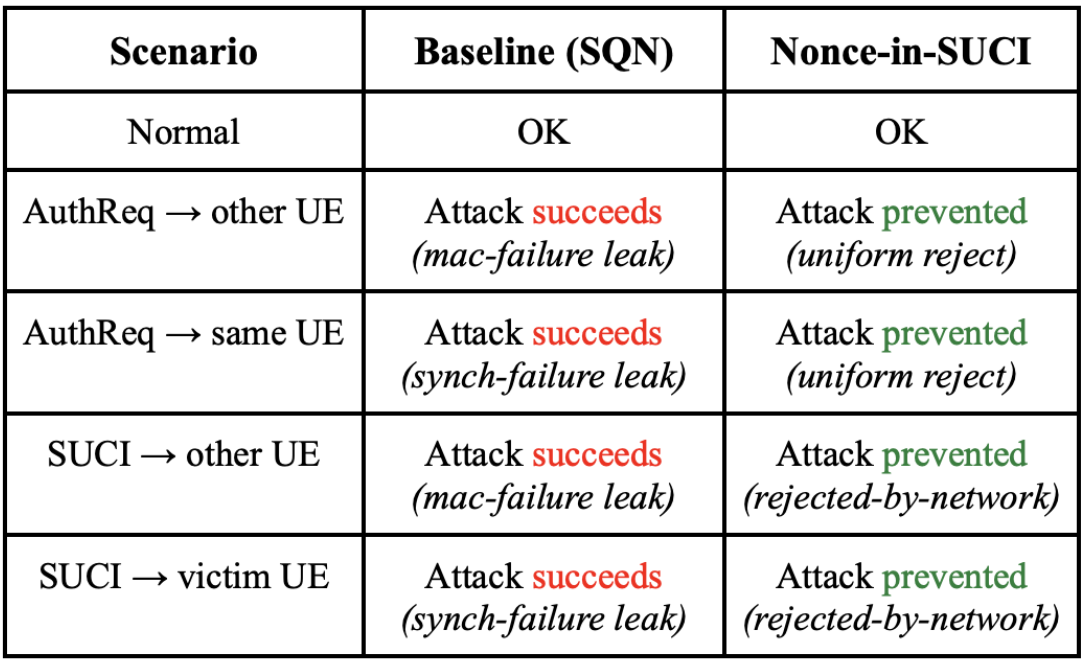}
  \caption{Replay attack outcomes for baseline (SQN) vs. nonce-in-SUCI. 
           The results are organized by scenario and highlight the contrast 
           between baseline and nonce-in-SUCI authentication behavior.}
           \vspace{-0.5cm}
  \label{tab:TableI}
\end{table}

\subsection{Testing Methodology and Results}
To evaluate the two variants, we designed controlled replay experiments that directly correspond to the vulnerabilities introduced in Section~III. We describe these results in detail below, with each scenario corresponding to the rows of Table~\ref{tab:TableI}.  

\subsubsection{Normal authentication}
As shown in the first row of Table~\ref{tab:TableI}, both the baseline and nonce-in-SUCI 
implementations complete the 5G AKA exchange successfully. The UE derives $\mathsf{RES}^*$ and the serving network 
verifies $\mathsf{HXRES}^*$. This outcome, labeled ``ok,'' denotes that authentication succeeded and the UE accepted $\mathsf{AUTN}$ as valid.  

\subsubsection{Replay of an Authentication Request}
Rows two and three of Table~\ref{tab:TableI} illustrate the effects of 
replaying a previously recorded attempt. Against a different UE, the baseline produces a 
\textit{mac-failure}, which occurs when the UE cannot verify the integrity check in $\mathsf{AUTN}$ because it does not 
possess the correct subscriber key. Against the same UE at a later time, the MAC verification succeeds (since the 
long-term key is correct), but the embedded sequence number falls outside the UE’s freshness window. The UE 
therefore produces a \textit{synch-failure} and returns an $\mathsf{AUTS}$ consisting of 
$(\mathsf{SQN} \oplus \mathsf{AK}^*) \parallel \mathsf{MAC\_S}$. Because these two error types are distinguishable in Table~\ref{tab:TableI},
an attacker can infer whether the subscriber is the same or different---the Failure Message Attack of Section~III-A.  
In contrast, the nonce-in-SUCI variant rejects both replays with the same generic response, thereby eliminating this vulnerability.  

\subsubsection{Replay of a SUCI}
Rows four and five of Table~\ref{tab:TableI} capture the effects of replaying a previously used SUCI. In the baseline protocol, the home network performs no replay protection: it decrypts the SUCI, assumes it is fresh, and issues a challenge for the subscriber. If this challenge is delivered to another UE, the outcome is a \textit{mac-failure}; if delivered to the original UE, the outcome is either ``ok'' (if the $\mathsf{SQN}$ is in-window) or a \textit{synch-failure} (if out of window). These differing outcomes leak whether the intended subscriber is present.  

In the nonce-in-SUCI variant, the network detects when a UE nonce has already been used. Such events are uniformly rejected and logged, preventing linkability while providing side information for higher-layer defenses. This does not fully eliminate the SUCI Replay Attack of Section~III-B but transforms it into harmless evidence rather than exploitable information.

In summary, these results demonstrate that our implementation directly mitigates the Failure Message Attack of Section~III-A by replacing distinguishable errors with a uniform reject outcome. It also begins to mitigate the SUCI Replay Attack of Section~III-B by detecting and logging nonce re-use, ensuring that replay attempts are harmless at the authentication layer while producing signals that higher-layer systems can use for further defense.

\balance

\vspace{-0.25cm}
\section{Conclusion}
\label{conclusion}
This paper makes three key contributions toward improving the privacy of the 5G Authentication and Key Agreement (AKA) protocol. First, a unified analysis of three known privacy-compromising attacks---the Failure Message Attack, the SUCI Replay Attack, and the AUTS Attack---is presented, framing them within a consistent notation and execution model to highlight their commonalities and differences. Second, five countermeasures are identified and evaluated, including both known techniques and novel refinements, analyzed under a shared framework that respects the constraints of the 5G system. Third, a complete Python implementation of the 5G AKA protocol is developed and used to reproduce the identified attacks and validate nonce-based mitigations through controlled replay testing.  

A key insight from this study is that deterministic sequence numbers (SQNs) are a major contributor to privacy weaknesses. Among the countermeasures considered, those that introduce a UE-generated random nonce---either within the SUCI or AUTS---emerge as the most robust and effective. The analysis shows that incorporating a UE nonce in SUCI ensures that all replay attempts are uniformly rejected. It is recommended that future communication systems, including privacy-sensitive deployments such as 3GPP-based military networks, adopt UE-generated nonces in lieu of or in addition to SQNs wherever unlinkability is a design goal.  

This enhancement can be implemented as a lightweight, proactive nonce challenge issued by the UE and carried as an optional NAS Information Element, preserving backward compatibility while adding negligible computational and bandwidth overhead. By combining a rigorous threat analysis with a minimal yet effective design change, the study contributes to a more privacy-resilient authentication architecture for 5G systems and beyond.  

Future work includes a full implementation and evaluation of the TS 33.501 resynchronization flow under real-world conditions, engagement with 3GPP working groups to propose standardization of the UE-initiated nonce as a new NAS IE, and deployment of these enhancements on a live 5G testbed (e.g., OpenAirInterface) to measure performance at scale. Additional research will also extend the adversary model to account for stronger threats such as compromised serving networks or side-channel adversaries, thereby further validating the security of the proposed solution.

% Don't forget to put back the acknowledgements

% \section*{Acknowledgments}
% This research was funded in part by the Department of Defense and by the National Science Foundation by way of the Center for Identification Technology Research (CITeR).

% \section{Appendix}
% \label{appendix}
% \input{sections/7_appendix.tex}


%%%%%%%%%%%%%%%%%%%%%%%%%%%%%%%%%%%%%%%%%%%%%%%%%%%%%%%%%%%%%%%%%%%%%%%%%%%%%%%%

\vspace{-0.25cm}
\bibliographystyle{IEEEtran}

\bibliography{library}

\end{document}